# Golden Vaterite as a Mesoscopic Metamaterial for Biophotonic Applications

*Roman E. Noskov\*, Andrey Machnev, Ivan I. Shishkin, Marina V. Novoselova, Alexey V. Gayer, Alexander A. Ezhov, Evgeny A. Shirshin, Sergey V. German, Ivan D. Rukhlenko, Simon Fleming, Boris N. Khlebtsov, Dmitry A. Gorin, and Pavel Ginzburg*

Dr. R. E. Noskov, A. Machnev, Dr. I. I. Shishkin, Prof. P. Ginzburg
Department of Electrical Engineering
Tel Aviv University
Ramat Aviv, Tel Aviv 69978, Israel,
E-mail: nanometa@gmail.com

Dr. R. E. Noskov, A. Machnev, Dr. I. I. Shishkin, Prof. P. Ginzburg
Light-Matter Interaction Centre
Tel Aviv University
Ramat Aviv, Tel Aviv 69978, Israel

Dr. I. I. Shishkin
Department of Physics and Engineering
ITMO University
Saint Petersburg 197101, Russia

Dr. M. V. Novoselova, Dr. S. V. German, Prof. D. A. Gorin
Center of Photonics & Quantum Materials, Skolkovo Institute of Science and Technology
Nobelya Str 3, Moscow 121205, Russia

Dr. S. V. German
Institute of Spectroscopy of the Russian Academy of Sciences
Troitsk 108840, Russia

Dr. A. V. Gayer, Dr. A. A. Ezhov, Dr. E. A. Shirshin
Faculty of Physics
M.V. Lomonosov Moscow State University
Leninskie Gory 1/2, Moscow 119991, Russia

Dr. A. A. Ezhov
Quantum Technologies Centre
M.V. Lomonosov Moscow State University
Leninskie Gory 1/2, Moscow 119991, Russia

Dr. A. A. Ezhov
A. V. Topchiev Institute of Petrochemical Synthesis of the Russian Academy of Sciences
Leninskii pr. 29, Moscow 119991 Russia

Dr. E. A. Shirshin
World-Class Research Center "Digital biodesign and personalized healthcare"
Sechenov First Moscow State Medical University
Trubetskaya 8-2, Moscow 119048, Russia

Dr. I. D. Rukhlenko, Prof. S. Fleming
The University of Sydney

Institute of Photonics and Optical Science
School of Physics
Camperdown, NSW 2006, Australia

Dr. I. D. Rukhlenko
Information Optical Technologies Centre
ITMO University
Saint Petersburg 197101, Russia

Prof. B. N. Khlebtsov
Lab of Nanobiotechnology
Institute of Biochemistry and Physiology of Plants and Microorganisms
Saratov 410049, Russia

Keywords: composite nanostructures; spherulites; Mie resonances; three-dimensional reconstruction; photoheating

Mesoscopic photonic systems with tailored optical responses have great potential to open new frontiers in implantable biomedical devices. However, biocompatibility is typically a problem, as engineering of optical properties often calls for using toxic compounds and chemicals, unsuitable for *in vivo* applications. Here, we demonstrate a unique approach to biofriendly delivery of optical resonances. We show that the controllable infusion of gold nanoseeds into polycrystalline submicron vaterite spherulites gives rise to a variety of electric and magnetic Mie resonances, producing a tuneable mesoscopic optical metamaterial. The three-dimensional reconstruction of the spherulites demonstrates the capability of controllable gold loading with volumetric filling factors exceeding 28%. Owing to the biocompatibility of the constitutive elements, golden vaterite paves the way to introduce designer-made Mie resonances to cutting-edge biophotonic applications. We exemplify this concept by showing efficient laser heating of gold-filled vaterite spherulites at red and near-infrared wavelengths, highly desirable in photothermal therapy and photoacoustic tomography.

## 1. Introduction

The fast-changing and evolving landscape of biomedical challenges motivates continued development of new functional platforms. Implantable devices made from biocompatible materials, capable of responding to optical signals and tailoring propagation of light waves can be employed in health monitoring applications and therapeutics[1]. Incorporation of such elements into living tissue can boost light-tissue interactions and shift conventional approaches towards precision medicine by opening new opportunities in sensing[2], photothermal therapy[3,4], photoacoustic tomography[5], and bioimaging[6]. One of the grand challenges on those pathways is the miniaturization of biocompatible photonic structures along with providing multifunctionality, such as monitoring of vital biological processes, light-responsive drug release, or local heating of a nanoscale area with the simultaneous measurement of its temperature.

The conventional approach towards enhanced light absorption and scattering in biosystems relies on using surface plasmon resonances of gold nanoparticles (Au NPs), providing nanoscale resolution and improved sensitivity. Variations in the shape and size of the NPs allows tuning their resonant frequencies across the visible and near-infrared spectral ranges, making Au NPs attractive as heating and passive contrast agents for biophotonic applications[4,6–8]. However, such structures typically support only one or two electric dipole resonances due to their deeply subwavelength dimensions. Meanwhile, providing on-demand multipole resonances in a nanoparticle may facilitate building up several modalities within a single platform.

An alternative route to the subwavelength optical localization is based on Mie resonances in high refractive index dielectric nanoparticles[9]. Isotropic and nonmagnetic structures support a variety of both electric and magnetic Mie modes, which occur when the incident light constructively interferes upon multiple reflections inside the particle. Hence, a ratio between a form factor and the light wavelength inside the particle remains nearly constant for a given eigenmode[9]. In practice, Mie resonances are very pronounced for nanoparticles of relatively high refractive index ($n > 2$), rendering efficient mode trapping into subwavelength volumes. This has led to the demonstration of many phenomena with practical outcomes, including Huygens-like nanoantennas, radiationless anapole modes, optical magnetic response, and many others[10]. In the visible and near-infrared spectral domains, the most commonly used high refractive index materials are titanium dioxide, germanium, and silicon[9]. Being favourable for nanoelectronics, semiconductor nanoparticles are almost never employed in biological and biomedical applications owing to their high toxicity and poor biocompatibility[11].

In contrast, biogenic materials typically possess quite low refractive indexes ($1.33 < n < 2$), unsuitable for generation of Mie resonances. Vaterite is an example of such a material which is actively used in bio-organic and nanomedical studies[12]. Being a metastable polymorph of calcium carbonate ($CaCO_3$), vaterite renders a variety of crucial functions in living organisms[13–17]. In nature, vaterite monocrystals can self-assemble in polycrystal micro- and nanoparticles, also referred to as spherulites. Due to the natural biocompatibility and strong porosity enabling high payload capacity, vaterite spherulites have found applications in targeted drug delivery and tissue engineering[18–23].

Here, we propose and demonstrate a new mesoscopic approach to on-demand engineering of electric and magnetic Mie resonances in submicron low-index vaterite spherulites by filling them with gold nanoinclusions. This makes possible to increase the effective permittivity of the structure and provide an efficient light trapping. The controllable variations in the volumetric gold filling factor allows tuning the spectral positions of Mie resonances from the dipole to octupole order within the visible and near-infrared ranges. A very good agreement between measured scattering and absorption properties of the gold-filled spherulites and the theoretical predictions validates the proposed approach. As an example of potential applications, we demonstrate efficient laser heating of *golden vaterite*. Both water suspensions and individual spherulites, being illuminated with red and near-infrared light, were investigated via time-resolved fluorometry and fluorescence-lifetime imaging microscopy (FLIM). The combination of various Mie resonances, high payload capacity for drugs, and the submicron size offer *golden vaterite* as a unique platform for multifunctional biomedical nanophotonic devices.

## 2. Results and discussion

### 2.1. Loading vaterite spherulites with gold nanoinclusions

Vaterite spherulites possess relatively high internal porosity, which can reach 40% of their overall volume, while pore sizes typically range between 30 and 50 nm[24]. In order to maximize spherulites' loading efficiency, we apply a state-of-the-art freezing-induced loading (FIL) technique[25]: controlled freezing forms ice lamellas which assist in pushing very small 3-nm Au NPs into the spherulite pores. Each loading cycle includes (i) injecting vaterite spherulite and gold-NP suspensions into a polymeric centrifuge tube and (ii) freezing the suspensions under gentle mixing, followed by (iii) thawing, (iv) centrifuging, and (v) washing the suspensions with pure water (Figure 1a). The desired loading concentration is achieved by varying the number of loading cycles, as detailed in Experimental Section.

We begin by characterizing the gold-loaded spherulites with high-resolution scanning electron microscopy (SEM) to reveal the efficiency of surface coating. The spherulites were imaged on a silicon substrate, with a high resolution (~0.7 nm), sufficient to distinguish the individual NPs (Figure 1b). In order to quantify the surface coverage efficiency versus the number of loading cycles, we introduce the surface coating factor, defined as the ratio of the area occupied by Au NPs to the spherulite's visible area. A backscattered electron detector was used to enhance contrast between vaterite and Au NPs (Experimental Section). The statistical analysis for a large number of SEM images shows that the mean surface coating factor monotonically grows from 0.168 for one loading cycle to 0.338 for seven cycles, as shown in Figure 1c and Figure S1.

To assess the efficiency of the volumetric loading, we sliced the samples using a focused ion beam (FIB) and then imaged their cross-sections with scanning transmission electron microscopy (STEM). The results for seven loading cycles are shown in Figure 1d and Figure S2. The internal vaterite pores are seen to be densely decorated with Au NPs, clearly demonstrating high volume loading in addition to surface coating. We also performed a three-dimensional reconstruction of Au NP distributions inside the spherulites by processing a z-stack of STEM images, which confirmed quite a dense filling (Figure 1e and Figure S4). The reconstruction made possible to precisely evaluate the distribution of Au NPs inside the spherulites. For the sake of clarity, we chose almost spherical particles, split them into many co-axial spherical layers, and calculated the local volumetric filling factor of each layer, as exemplified in Figure 1f and Figure S4. One can see that the concentration of Au NPs is the lowest in the spherulite centre, while its maxima are reached close to the spherulite surface and at $r \approx R/2$, where $R$ is the spherulite radius. This situation is typical for all particles analysed. Such spatial distribution fully meets expectations from the FIL technique, which is focused on pushing Au NPs from the spherulite surface to the centre. The additional local maximum around $r \approx R/2$ can be attributed to the presence of relatively large central cavities in the initial spherulites, whose surfaces efficiently absorb Au NPs (Figure 1d). Interestingly, similar radially inhomogeneous internal structure of silica spheres was used to manipulate stop-bands in photonic crystals[26,27].

The total filling factor of a single spherulite is given by $f_0 = V^{-1} \sum_{n=1}^{N} v_n f_n$, where $V$ is the spherulite volume, $N$ is the number of layers, and $v_n$ and $f_n$ are the volume and the filling factor for the $n$th layer. For seven loading cycles the mean of $f_0$ for 10 spherulites analysed was found to be 0.28, in a good agreement with the evaluation of the surface coating factor and the measurements of the scattering and absorption spectra discussed below.

It is important to note that when the size of metallic nanoparticles drops below the electron mean free path (~10 nm for gold), their optical response does not obey the Drude model, and the quantum approaches should be used instead[28,29]. This, however, is not the case for our structures since 3-nm Au NPs get stuck together during the loading process and form aggregates, exceeding 10 nm, as evidenced by Figures 1b, 1d, and 1e.

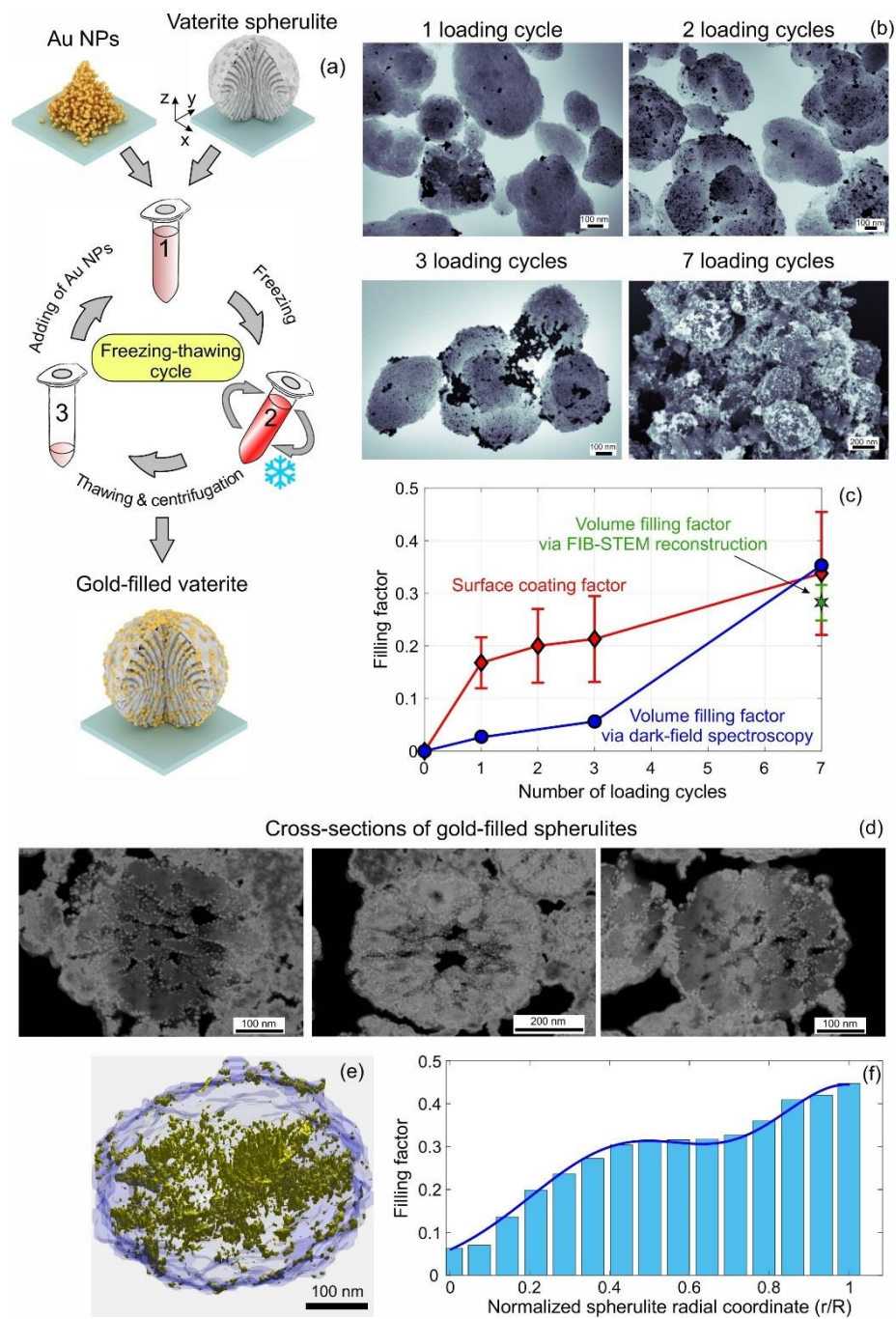

**Figure 1.** (a) Schematics of infusing gold nanoparticles (Au NPs) into vaterite spherulites via freezing-induced loading and (b)–(f) characterization of the resulting gold-filled spherulites. The variations in the number of loading cycles allows control of the filling factor. SEM micrographs for one, two, and three loading cycles in (b) are presented in the inverse color scheme to highlight the sparse distributions of gold nanoinclusions (black color) whereas the case of seven loading cycles is shown in the normal color scheme. Panel (c) shows the surface coating and volume filling factors as functions of the number of loading cycles. The mean and the standard deviation are obtained by the statistical analysis of a large number of spherulites in the high-resolution SEM and STEM images (Supporting Information). Also, the volume filling factor is evaluated by numerical fitting to the dark-field scattering spectra shown in Figure 2. The cross-sections of vaterite spherulites loaded with 7 cycles in (d) are obtained by etching with the focused ion beam and then imaged via the scanning transmission electron microscopy. SEM and STEM images were taken before annealing. The 3D reconstruction in (e) is performed by processing a z-stack of STEM images in MATLAB. Panel (f) shows spatial distribution for Au NPs from the reconstruction in (e). The distribution is fitted by a sum of two Gaussian functions (Supporting Information).

Our technique for the vaterite synthesis allows variations in the spherulite size from 300 nm up to 1000 nm. However, in what follows we focus on 400-nm, 500-nm, and 800-nm spherulites, which demonstrate the largest number of Mie resonances in the visible and near-infrared (NIR) ranges.

## 2.2. Annealing

The FIB-STEM analysis showed that the characteristic scale of the surface roughness for gold aggregates inside the vaterite spherulites is comparable with the electron mean free path, reaching ~10 nm. Such plasmonic structures have additional optical losses associated with significant electron scattering on metallic surface grains[30]. Phenomenologically, this effect can be described as an additional loss factor $\zeta$ to the standard damping rate $\gamma$ in the Drude equation for the gold permittivity as

$$\varepsilon_{\mathrm{Au}} = 1 - \frac{\omega_p^2}{\omega^2 + i\zeta\gamma\omega}, \tag{1}$$

where $\omega_p$ is the plasma frequency, and $\zeta$ quantifies the difference between the damping term in bulk gold and a considered structure. In what follows we use the gold Drude parameters from Ref.[31].

In order to reduce the additional scattering losses, we annealed the samples at 300 °C for 2 min in a thermal processing system (Memmert UN30) by following the protocol suggested in Ref.[30]. While being high enough to reduce the roughness of the gold aggregates, this temperature is still insufficient to drive a phase transition of vaterite to calcite[24]. The correction factor $\zeta$ was obtained by fitting the measured scattering and absorption spectra with the results of numerical simulations. Without annealing, the samples with any number of loading cycles demonstrated $\zeta \approx 5$ (not shown).

As it will be shown in Sections 2.3 and 2.4, the annealing had best performance for the structures with the highest filling factors. Specifically, for the particles with 1 and 3 loading cycles $\zeta$ dropped to as low as 3; while for the particles with 7 loading cycles annealing resulted in $\zeta = 1$. This effect can be attributed to the dependence of the characteristic size of the gold aggregates inside the spherulites on the number of loading cycles. For small loadings, this size is about 10 nm (Figure 1b), which means that the surface electron scattering is significant regardless the surface roughness. The increase of this size with the filling factor gives rise to a more pronounced contribution of the surface roughness to $\zeta$, which is reduced by annealing.

## 2.3. Dark-field spectroscopy

The scattering properties of gold-filled spherulites were characterized using dark-field spectroscopy (Experimental Section), which allows spectral analysis of subwavelength particles by eliminating the direct light via Fourier filtering (Figure 2a). Figures 2b–2d compare the measured and the simulated dark-field scattering spectra of empty and gold-filled vaterite spherulites obtained with different numbers of loading cycles. In order to evaluate the exact shapes and sizes of the optically characterized spherulites, we imaged them with SEM. For each set of samples, we recorded dark-field scattering spectra for about 30 particles and selected the spectra of spherical-like particles with diameters close to 500 nm.

The scattering spectra of empty spherulites are free of any notable features, because small-size low refractive index particles do not resonate (Figure 2b). The first loading cycle produces a clear maximum around 610 nm, which can be reproduced theoretically with a gold volumetric filling factor of 0.02 and $\zeta = 3$. A further increase in the number of loading cycles gradually raises the filling factor, redshifts the scattering maximum to 700 nm, and results in the appearance of the additional scattering maxima around 530 nm for seven loadings with $f = 0.36$ and $\zeta = 1$. These results correlate well with the evaluations of the surface coating factor from the high-resolution SEM micrographs and the volume filling factor from the 3D reconstruction in Figure 1c.

It is important to note that the scattering intensity for all measured samples drops to zero around 800 nm while simulations predict a finite signal in this region. This discrepancy is caused by the fact that the broadband antireflection coating for the optical components of the dark-field microscope works only in the region 400-750 nm. This also explains why the third scattering maximum at 755 nm in Figure 2e, predicted by numerical simulations, is not observed in the experiment.

## 2.4. Absorption measurements

The absorption spectra of individual spherulites were measured using a custom-built experimental setup detailed in Experimental Section and shown in Figure S5. The transmission spectra of the spherulites in the wavelength range of 500-800 nm were measured after imaging of the substrate performed by scanning microscopy which detected the transmitted light in the whole spectral range. A receiving objective (Figure 2f) and the confocal aperture size allowed to obtain spatial resolution of 1 μm, which was sufficient to distinguish between single submicron spherulites in diluted samples. Next, we selected the regions of interest, whose size being approximately equal to one micron in the entire spectral range and measured transmission spectra of the spherical-like particles with sizes close to 500 nm relative to the adjacent areas free from particles.

Figures 2g and 2h show the results for the spherulites with three and seven loading cycles. The three loading cycles led to a clear maximum around 550 nm along with a gradual decreasing for longer wavelengths. This behaviour can be reproduced theoretically with a gold volumetric filling factor of 0.08 and $\zeta = 3$. Seven loading cycles give rise to three distinct absorption maxima at 630, 730, and 780 nm corresponding to $f = 0.32$ and $\zeta = 1$. These results are also in a good agreement with the straightforward material characterization presented in Figure 1c.

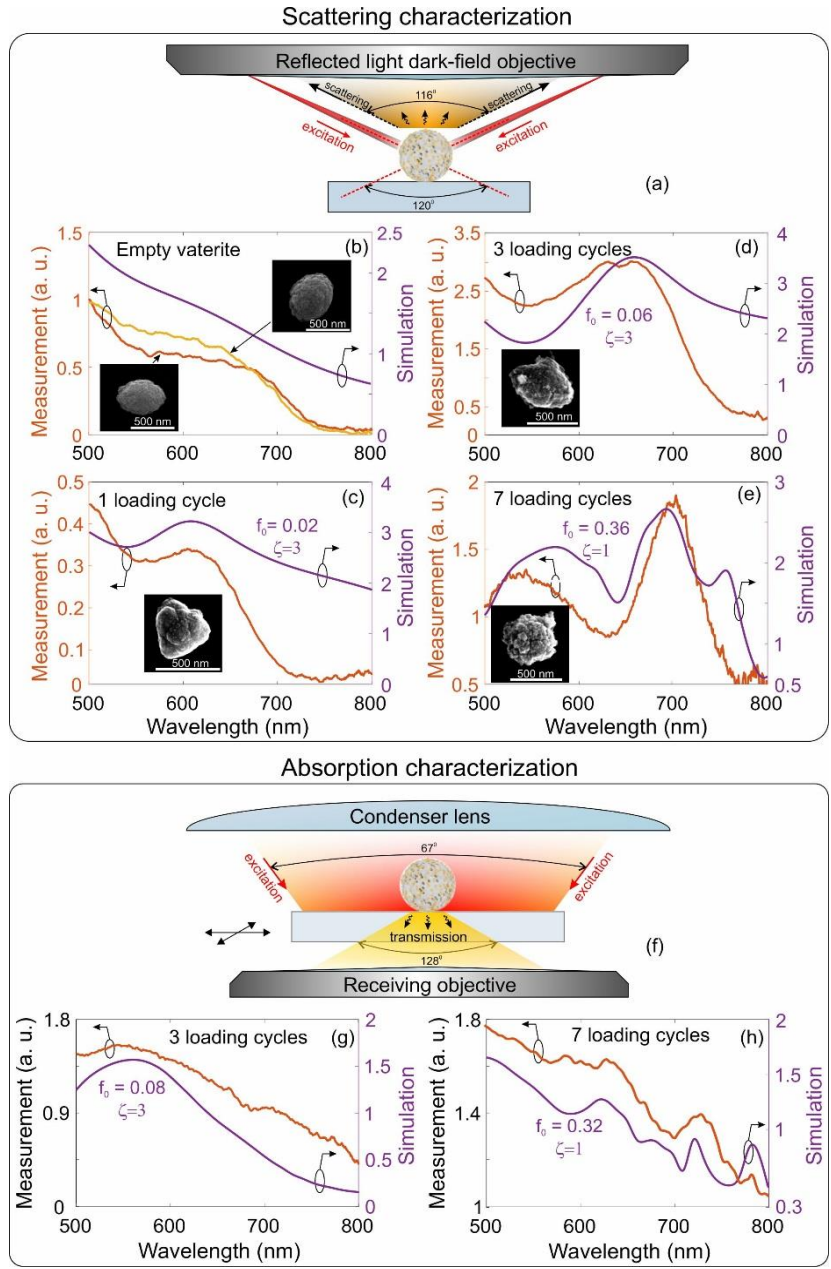

**Figure 2.** Characterization of scattering and absorption properties of annealed gold-filled vaterite spherulites. (a) Schematic of the dark-field spectroscopy setup; (b)-(e) measured and simulated dark-field scattering spectra for empty and gold-filled vaterite spherulites obtained with different

numbers of loading cycles; (f) schematic of the setup for the measurement of the spherulites'
absorption; (g) and (h) measured and simulated absorption spectra for gold-filled vaterite
spherulites. The simulation axis denotes the dark-field scattering and the absorption cross-
sections normalized by the spherulite geometrical cross-section. $f_0$ is the volume filling factor
of Au NPs and $\zeta$ is the additional loss factor in the Drude model used in the simulations to fit
in the measurements. All simulations have been performed for a 500-nm spherulite.

## 2.5. Numerical model

A polycrystal vaterite spherulite can be represented by the so-called 'dumbbell' or 'heap of
wheat' model[32] as a bundle of fibers tied together at the center and spread out at the ends.
Each fiber consists of small (about 30 nm in size) positive uniaxial monocrystals with $\varepsilon_o \approx$
$2.4$ and $\varepsilon_{eo} \approx 2.72$, aligned to the tangents of the fiber. The distribution of the unity optical
axes in the spherulite is well approximated by the family of co-focused hyperboles
symmetrically rotated with respect to the $z$-axis[32,33] (Figure 1a). In what follows we assume
that the focal position of the hyperboles is half of the radius of the sphere that fits submicron
vaterite spherulites well.

With these considerations, the effective permittivity of the vaterite scaffold is described
by a symmetric non-diagonal tensor whose components can be obtained by translating the
permittivity tensors of individual monocrystal subunits from the local coordinate systems,
associated with their unity optical axes, into the global coordinate one, anchored to the
spherulite center[32]

$$\hat{\varepsilon}_{\text{vat}} = \begin{pmatrix} \varepsilon_{xx} & \varepsilon_{xy} & \varepsilon_{xz} \\ \varepsilon_{xy} & \varepsilon_{yy} & \varepsilon_{yz} \\ \varepsilon_{xz} & \varepsilon_{yz} & \varepsilon_{zz} \end{pmatrix}. \tag{2}$$

The coordinate dependences of the tensor components can be found in Supporting Information.

Deeply subwavelength size of Au NPs allows us to treat the electromagnetic response
of the gold-filled vaterite spherulites within the Maxwell Garnett formalism, which introduces
the effective permittivity tensor as[34]

$$\hat{\varepsilon}_{\text{eff}} = \hat{\varepsilon}_{\text{vat}} + f(\varepsilon_{\text{Au}}\hat{I} - \hat{\varepsilon}_{\text{vat}})\left(\hat{\varepsilon}_{\text{vat}} + \frac{1-f}{3}(\varepsilon_{\text{Au}}\hat{I} - \hat{\varepsilon}_{\text{vat}})\right)^{-1}\hat{\varepsilon}_{\text{vat}}, \qquad (3)$$

where $\hat{I}$ is the identity matrix, and $f$ is the volume filling factor of gold. The spatial inhomogeneity of Au NPs distribution inside the spherulites can be included into the model via the radius-depended filling factor as follows,

$$f(r) = \frac{Vf_0\rho(r)}{4\pi\int_0^R \rho(r')r'^2 dr'}.$$

Here $\rho(r) = A_1\exp\left[-\frac{(r-\Delta r_1)^2}{2\delta_1^2}\right] + A_2\exp\left[-\frac{(r-\Delta r_2)^2}{2\delta_2^2}\right]$ is the function fitting the data revealed by the FIB-STEM analysis with adjustable parameters $A_{1,2}$, $\Delta r_{1,2}$, and $\delta_{1,2}$ (Figure 1f, Figure S4, and Table S1). We applied normalization in the model using the same total filling factor $f_0$ as that extracted by the FIB-STEM analysis. In further calculations we used $\rho(r)$ shown in Figure 1f.

It is instructive to note that the surface coating of the samples by Au NPs did not exceed the typical surface roughness of spherulites of 20-30 nm. Therefore, our model takes into account the presence of Au NPs on the spherulite's surface via the dependence of the effective spherulite permittivity on the radial coordinate.

In order to get physical insights into light scattering by the gold-filled spherulites, we use the Cartesian multipole expansion of the scattered field. This presentation explicitly shows the contributions of the high-order irreducible tensors (basic multipoles) to the scattering cross-section. The multipole expansion of the scattering cross-section up to the terms of the third order is given by[35,36]

$$\sigma_{\text{scat}} = \frac{k^4}{6\pi\varepsilon_0^2|\mathbf{E}_0|^2}\left|p_j\right|^2 + \frac{k^4\varepsilon_h}{6\pi\varepsilon_0^2|\mathbf{E}_0|^2}\left|m_j\right|^2 + \frac{k^6\varepsilon_h}{80\pi\varepsilon_0^2|\mathbf{E}_0|^2}\left|Q_{jk}^{(e)}\right|^2$$
$$+ \frac{k^6\varepsilon_h^2}{80\pi\varepsilon_0^2c^2|\mathbf{E}_0|^2}\left|Q_{jk}^{(m)}\right|^2 + \frac{k^8\varepsilon_h^2}{1890\pi\varepsilon_0^2|\mathbf{E}_0|^2}\left|O_{jkl}^{(e)}\right|^2 + \frac{k^8\varepsilon_h^3}{1890\pi\varepsilon_0^2c^2|\mathbf{E}_0|^2}\left|O_{jkl}^{(m)}\right|^2, \qquad (4)$$

where Einstein's summation notation is used, $k$ is the vacuum wave number, $|\mathbf{E}_0|$ is the amplitude of the incident field, $\varepsilon_h$ is the permittivity of the host media, $\varepsilon_0$ is the permittivity of vacuum, $p_j$ and $m_j$ are the basic electric dipole (ED) and magnetic dipole (MD) moments, $Q_{jk}^{(e)}$ and $Q_{jk}^{(m)}$ are the basic electric and magnetic quadrupoles (EQ and MQ), $O_{jkl}^{(e)}$ and $O_{jkl}^{(m)}$ are basic electric and magnetic octupoles (EO and MO). The exact expressions for them can be found in Refs.[35,36].

We next apply the finite element method (FEM) by COMSOL Multiphysics to solve the problem of light scattering by the gold-filled vaterite spherulites. The scattering cross-section is then calculated with two methods: following Equation (4) and by the straightforward integration of the Poynting vector. The proximity of the thus obtained two cross-section spectra evidences the accuracy of the above expansion.

When simulating the dark-field spectroscopic experiment and the absorption measurements (Figure 2), we take into account all the features of the setups, including the angle of illumination, the refractive index of the glass substrate, and the numerical aperture (NA) of the collection optics. Since we use unpolarized light for samples excitation and do not control the orientation of the spherulites, we calculate the mean of the four scattering spectra obtained for the TE and TM polarizations of the incident wave and when the major optical axis of the spherulite is either parallel or orthogonal to the air–substrate interface.

## 2.6. Theoretical analysis of scattering and absorption

In order to characterize the optical properties of gold-filled vaterite spherulites, we calculate their scattering and absorption efficiencies (defined as the scattering and absorption cross-sections normalized by the particle geometric cross-section) and magnetic field enhancement in the visible and NIR regions for gold filling factors from 0 to 0.32. The results for 400-nm and 500-nm spherulites in vacuum are summarized in Figure 3. The empty spherulites have vanishing absorption due to the high optical transparency of vaterite (Figures 3a and 3b). The broad maxima in the scattering efficiency at the short wavelengths comparable to the particle sizes are caused by the low-Q-factor MD, MQ, and MO resonances due to the low refractive index of vaterite.

Loading spherulites with gold dramatically changes their optical properties. Small filling factors ($f \sim 0.01$) result in a well-pronounced absorption peak around 540 nm due to the ED plasmon resonance of individual Au NPs and the optical transparency of vaterite. For such small concentrations, Au NPs are electromagnetically uncoupled and can be considered independently. The localized-plasmon peak splits into several resonances when the filling factor exceeds 0.08, because gold-filled spherulites turn into metamaterials and their effective permittivities (Equation 3) start governing the light–matter interaction. In this case, the scattering spectrum of the spherulites consists of the overlapping collective Mie modes even for very small filling factors while the ED scattering peak of individual Au NPs disappears, manifesting the formation of a metamaterial. A further increase in the filling factor redshifts Mie resonances and enhances their Q-factors.

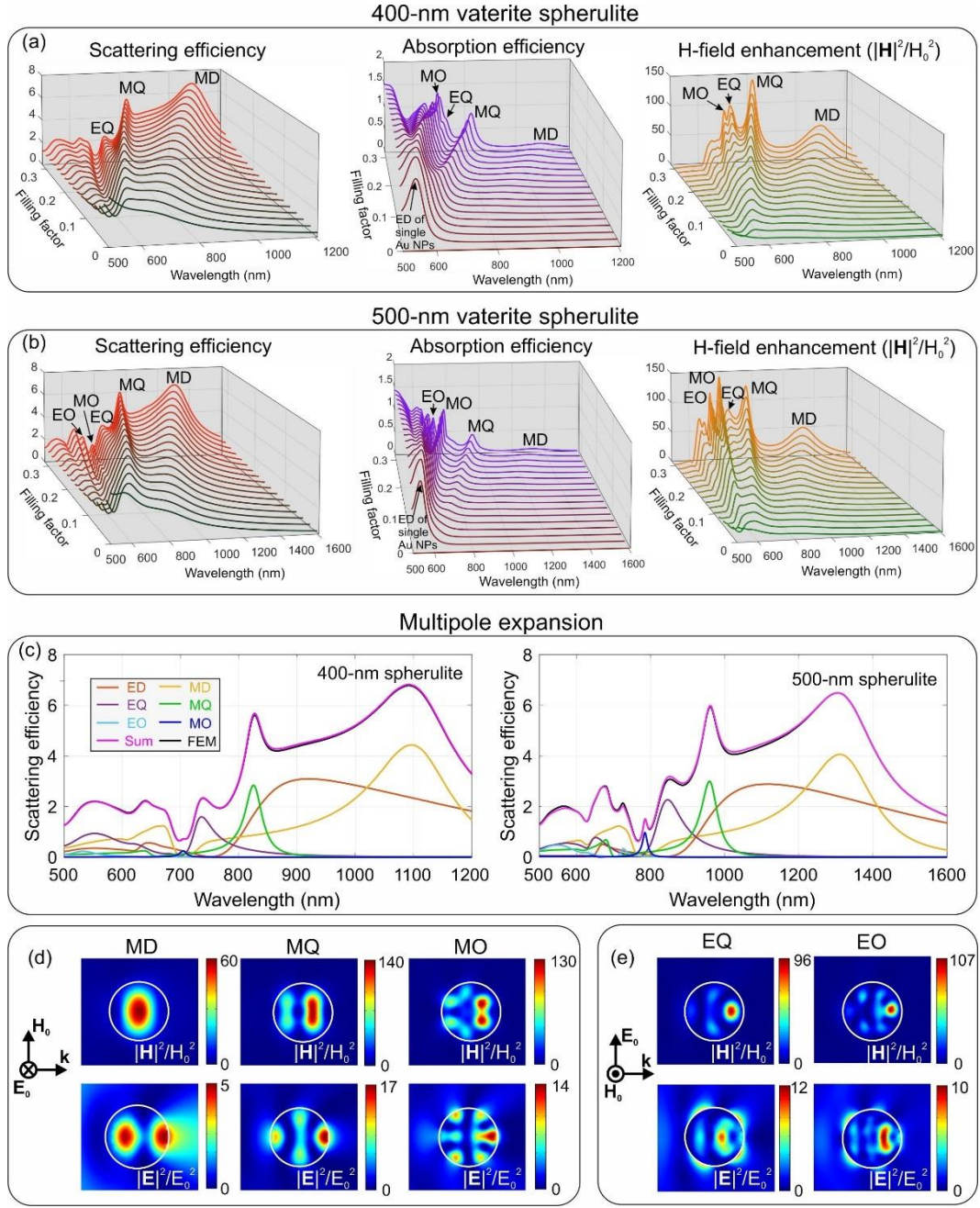

**Figure 3.** Scattering efficiency, absorption efficiency, and maximum local magnetic field enhancement for (a) 400-nm and (b) 500-nm gold-filled vaterite spherulites. The scattering and absorption efficiencies are the cross-section spectra normalized by the spherulite geometric cross-section; the filling factor of gold varies from 0 to 0.32 with 0.02 steps. Panel (c) shows characterization of Mie resonances by the Cartesian multipole expansion of scattering cross sections for 400-nm and 500-nm gold-filled vaterite spherulites. The convergence of the multipole contributions sum (purple) to the scattering efficiency obtained by the numerical integration of the Poynting vector (black) indicates a high accuracy of the adopted multipole

expansion. Panels (d) and (e) show spatial distributions of magnetic and electric fields of various Mie resonances for a 500-nm spherulite and $f_0 = 0.32$: the MD, MQ, EQ, MO, and EO modes are shown for the wavelengths of 1310 nm, 960 nm, 845 nm, 785 nm, and 726 nm, respectively. In all cases, the major optical axis of the spherulite is aligned with the external electric field polarization and $\zeta = 1$.

For sufficiently large filling factors of Au NPs five sharp Mie resonances — MD, EQ, MQ, EO, and MO — are clearly seen in the scattering spectrum of the spherulites (Figure 3c) and visualized in Figures 3d and 3e. The ED resonance is indistinguishable due to its low Q-factor. Because the onset of interband electron transitions makes gold to be highly absorptive below 500 nm, there are no higher-order Mie resonances at short wavelengths. All of these resonances are accompanied by a strong enhancement of the magnetic and electric field, with enhancement factor squared reaching 140 and 17, respectively, similar to the case of Mie resonances in high-index particles[9].

The Cartesian multipole expansion also enables the peaks observed in the scattering and absorption spectra to be associated with the type of Mie resonance. Specifically, the scattering maxima in Figures 2c and 2d for the samples with 1 and 3 loading cycles correspond to the MQ resonance (Figure 3b). The absorption spectra for the sample with 3 loading cycles shows a broad maximum corresponding to the ED resonances of individual Au NPs (Figure 2g). For the spherulite with 7 loading cycles (Figure 2e) the longest wavelength scattering maxima (predicted only by simulations) is caused by the EO mode, while other two maxima observed stem from the constructive interference of four multipoles: ED+MD+MQ+EQ for the peak at 700 nm and MD+MQ+EQ+EO for the peak at 530 nm. The absorption maxima for a similar spherulite with 7 loading cycles, located around 785 nm and 730 nm (Figure 2h), are due to the MO and EO resonances, respectively.

To conclude, the filling factor of Au NPs is a core parameter that allows engineering a wide variety of Mie resonances in vaterite spherulites, covering a broadband range from 500 nm to 1300 nm. Hence, gold-filled vaterite spherulites markedly differ from Au NPs extensively employed in biomedical applications which typically demonstrate only one or a few ED resonances tuneable by changes in the NP geometrical parameters[6–8]. It is instructive to note that a geometrical degree of freedom in adjusting the resonance properties can be employed for vaterite spherulites as well. As we have shown recently[12], the shape of submicron vaterite particles can be controllably tuned in a wide range from torus to prolate ellipsoids by modifying the synthesis protocol. This study will be published elsewhere.

## 2.7. Laser-induced heating of gold-filled vaterite spherulites

To demonstrate the great potential of gold-filled vaterite spherulites for biophotonic applications, we studied the thermometry of spherulites suspended in water under red and infrared laser radiation using time-resolved fluorometry and probed the temperature of single particles by FLIM. Both approaches are based on the association of the temperature rise with the decrease in the fluorescence lifetime of the probe (dye). To this end, the gold-filled vaterite spherulites were additionally loaded with a conjugate of bovine serum albumin (BSA) and Cy5 dye using the FIL technique, as detailed in Experimental Section. The quantity of the conjugate per single spherulite was controlled to be much less than the quantity of Au NPs (~0.02 pg vs ~0.3 pg) to exclude changes in the optical properties of the samples due to the dye. Accounting for quite inertial cooling of the samples (ca. 5 min, Figure 4a), we shifted in time the laser heating of the samples and the fluorescence lifetime-based measurements to provide high signal-to-noise ratio (see Experimental Section).

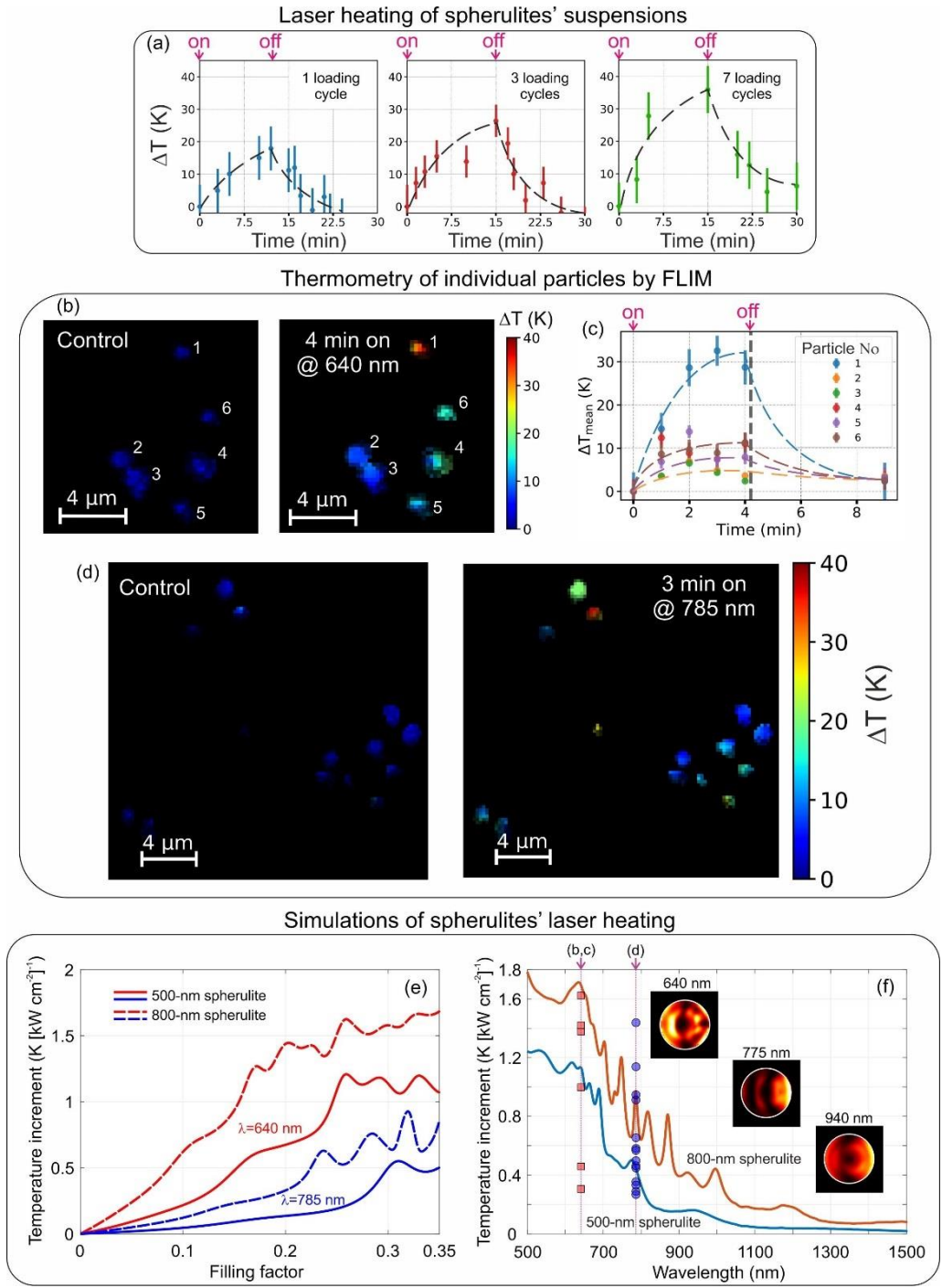

**Figure 4**. Characterization of laser-induced heating of gold-filled vaterite spherulites. (a) Temperature kinetics for aqueous solutions of spherulites with 1, 3, and 7 loading cycles measured by time-resolved fluorometry. Purple arrows indicate the times of laser turning on and off. The concentration of spherulites in all samples was $2 \times 10^7$ ml$^{-1}$, and heating was performed at the wavelength of 660 nm with the laser intensity of 0.1 kW cm$^{-2}$. The temperature of individual spherulites with 7 loading cycles was probed by FLIM for pumping at [(b) and (c)] the wavelength 640 nm and the laser intensity of 8 kW cm$^{-2}$ and (d) the wavelength of 785 nm and the laser intensity of 15 kW cm$^{-2}$. Dashed curves in (a) and (c) mark the exponential fits as guides for the eye to demonstrate the heating and cooling regimes. (e) Temperature increment as a function of the filling factor calculated for 500-nm and 800-nm spherulites heated at the wavelengths of 640 nm (red) and 785 nm (blue). (f) The temperature increment vs the wavelength for the 500-nm and 800-nm spherulites with $f_0 = 0.32$. Insets show distributions of the optical power loss density at the wavelengths corresponding to Mie resonances for a 500-nm particle. Red squares and blue circles mark the steady-state temperature increment for the particles from (b) and (d), respectively. The purple arrows indicate the pumping wavelengths for (b,c) and (d). In all simulations $\zeta = 1$.

First, we measured calibration curves linking the temperature and the fluorescence lifetime for the aqueous solutions of the gold-filled vaterite spherulites (Figure S6). Then, we irradiated the aqueous solutions of gold-filled spherulites with a 660-nm light of intensity 0.1 kW cm$^{-2}$ and recorded the temporal kinetics of the temperature (Figure 4a). The dye excitation was performed by a 635-nm laser diode, and emission was detected through a 660-nm long-pass filter. In full agreement with the analysis of the spherulites absorption (Figure 3), one may observe that an increase in a number of loading cycles associated with larger filling factors leads to stronger heating.

Finally, we probed the temperature kinetics for individual spherulites with 7 loading cycles immobilized on an interface between water and a glass plate using FLIM. The dye excitation was performed at 640 nm, and the fluorescence signal was detected through the 655-725 nm bandpass filter (fitting well the fluorescence window for the Cy5 dye). Figures 4b and 4c show the results for heating at the wavelength of 640 nm and the intensity of 8 kW cm$^{-2}$, and Figure 6d shows the results for heating at the wavelength of 785 nm and the intensity of 15 kW cm$^{-2}$.

In all cases, the particles were well heated with a saturation temperature rise varying from a few K to 40 K. To gain a deeper insight into these results, we performed simulations in COMSOL Multiphysics using the numerical model of the spherulites described above. Figure 4e shows the simulated temperature increment as a function of the filling factor for the wavelengths of 640 nm and 785 nm used in the FLIM measurements. We considered 500-nm and 800-nm spherulites within a full range of gold loadings observed in the material characterization (Figure 1). The temperature increment is seen to grow with the filling factor, and local maxima correspond to Mie resonances.

We also considered the spectrum of the temperature increment for 500-nm and 800-nm spherulites with $f_0 = 0.32$, shown in Figure 4f. Mie resonances, manifested as local maxima, are visualized in the insets as spatial distributions of the optical power loss density which acts as a source of heating. Specifically, the resonances at 940 nm and 775 nm for a 500-nm particle correspond to the MQ and MO modes, respectively, while the rest of the Mie modes are of higher orders. Interestingly, the inhomogeneous distribution of Au NPs inside the spherulite appears as the minima of the optical power loss density in the centres of spherulites for all Mie modes.

To compare theoretical predictions with the results of FLIM, we translated the observed mean temperatures of the particles (Figures 4b and 4d) and the driving optical intensities into the temperature increments and marked them in Figure 4f as red squares for pumping at 640 nm and blue circles for pumping at 785 nm. A very good agreement between measurements and simulations can be seen. The different temperature rise for various particles can be attributed to the variations in the filling factor and the particle size. Also noteworthy is that some particles were heated much more strongly than predicted by theory, which indicates the presence of particle clusters that absorb light better than single spherulites.

Thus, gold-filled spherulites diluted in water can be efficient subwavelength heating agents absorbing light by Mie resonances in the broadband domain from 500 to 1300 nm. This paves the way for using gold-filled vaterite spherulites in photothermal therapy[3,4] and photoacoustic tomography[5].

## 3. Conclusion and outlook

We have proposed and demonstrated a novel biocompatible platform for flexible engineering of lower and higher order Mie resonances. Our concept is based on modifying low-index porous vaterite spherulites with gold nanoinclusions using the state-of-the-art freezing-induced loading technique. This new designer metamaterial approach allows tuning the electromagnetic response of the spherulites by controlling the concentration of gold inside them.

Our findings suggest introducing Mie resonances into implantable photonic components with the help of low-index biocompatible vaterite spherulites. The extremely wideband tunability of Mie resonances in *golden vaterite*, covering the first (700–1000 nm) and second (1000–1500 nm) biologically relevant transparency windows, along with high payload capacity which allows filling them with Au NPs, pharmacological agents, and fluorescent tags, opens promising prospects for multifunctionality and enables the use of a single structure for sensing[2], photothermal therapy[4], photoacoustic tomography[5], bioimaging[6], and targeted drug delivery[19,21]. Further to that, gold-filled vaterite spherulites can serve as basic elements of nanophotonic circuitry in implantable biophotonic devices[1].

## 4. Experimental Section
### 4.1. Samples preparation

**Preparation of Au NPs.** Au NPs were synthesized by the protocol reported by Duff *et al*.[37] as follows: 0.5 mL of 1M NaOH and 12 µL of tetrakis(hydroxymethyl)phosphonium chloride were added to 9 mL of water under vigorous stirring. Then 0.5 mL of 8% $HAuCl_4$ were added quickly to the stirred solution. The reaction mixture turned from colorless to dark brown, which indicated the formation of gold nanoparticles.

**Preparation of $CaCO_3$ particles.** Submicron vaterite templates with the targeted size of about 500 nm were synthesized as described by Novoselova *et al*.[38] In brief, 4 g of glycerol were mixed with 0.4 mL of $CaCl_2$ and 0.4 mL of $Na_2CO_3$ aqueous solutions of equal concentrations (0.5 M) under vigorous stirring at 700 rpm at room temperature. In 60 min, the solution became cloudy, indicating the precipitation of $CaCO_3$. Then, the suspension was centrifuged, and the precipitate was washed with deionized water several times to remove glycerol.

**Loading of Au NPs.** The loading of vaterite spherulites with Au NPs was performed by the freezing-induced loading approach[25] as follows: 1 mL of colloidal Au NPs was added to the suspension of vaterite spherulites. A microcentrifuge tube with the reaction mixture was kept in a freezing chamber at −20°C for 2 h and stirred slowly and constantly using a rotator TetraQuant R1. After that, the samples were thawed at room temperature and washed. For washing the particles, they were separated by centrifugation at 8000 rpm for 1 min. The loading cycles were increased from one up to seven to get different concentrations of the loaded Au NPs. The supernatant was then analyzed for the Au NPs concentration. The concentration of Au NPs was determined by measuring the absorbance at 400 nm using the Infinite M Nano+ (Tecan Group Ltd., Switzerland) as described in Hendel *et al.*[39]

**Loading of Cy5 dye.** Cyanine 5–N-hydroxysuccinimide ester (Cy5–NHS) was purchased from Lumiprobe (Hannover, Germany). Cy5-labelled bovine serum albumin (BSA) was prepared by a standard Lumiprobe protocol and was purified by dialysis. Gold-filled vaterite spherulites were loaded with 1 ml of a conjugate of BSA and Cy5 (2 mg ml$^{-1}$) by the FIL method. Then the particles' suspension was centrifuged and washed twice with pure water. The amount of Cy5 per particle was determined via supernatants using a fluorescence spectrophotometer Infinite® M Nano (TECAN).  To this end, the supernatants were collected after each washing step during the particle loading. The emission spectra of the particle supernatants were registered at the excitation wavelength of 650 nm. The average content of Cy5 per single spherulite was found to be 0.022, 0.021, and 0.019 pg for the samples with 1, 3, and 7 loading cycles, respectively.

### 4.2. Spherulites characterization by SEM and STEM.

The SEM images in Figure 1b were taken in the backscattered electron mode by the high-resolution microscope Zeiss Gemini SEM 300 with an acceleration voltage of 5 kV at 0°C and under the pressure of $7.81 \times 10^{-7}$ mbar. The slicing of the spherulites with FIB and imaging their cross-sections by STEM in Figure 1d were performed with a Thermo Scientific Helios 5UC system.

## 4.3. Dark-field spectroscopy

The scattering spectra were acquired with the Zeiss Axiolab 5 V1 microscope equipped by an Zeiss EC Epiplan 100x/0.85 HD objective where an eyepiece was replaced by a port adapter for Zeiss Axioscop Miscroscopes (Thorlabs) in order to collect and guide the scattered field to the fiber-optic spectrometer (Avantes AvaSpec-UL2048L) via a multimode fiber. The influence of the substrate was excluded by subtracting the background signal, collected in the close vicinity of the analyzed particle, from the scattering spectrum of the particle. The resulting spectra were normalized by the reference signal, which was collected by replacing the sample with a spectralon target. The collection angle was set to be 116°.

In order to evaluate the exact shapes and sizes of the optically characterized spherulites, we imaged them with a FEI Quanta 200 SEM microscope. The spherulites were coated with chromium prior to imaging, and the samples were identified using alignment marks on the substrate.

## 4.4. Absorption spectroscopy

The absorption spectra of individual particles were measured using a custom-built experimental setup based on a FluoView FV1000 confocal laser scanning microscope (CLSM) mounted on an IX81 motorized inverted microscope and fiber optic spectrometer Ocean Optics QE Pro (Figure S5). Optem VSI 220 fiber optic illuminator equipped with 150 W halogen lamp was used as a light source. The filter built into the light source was removed to increase the spectral range. An additional spectrum correction filter was used to ensure that the magnitude of the spectrometer signal in the entire spectral range was within the dynamic range. The light was supplied to a condenser through a light guide, and the geometry of the transmitted light was used. Dichroic filters were removed from the optical path. The removable mirror was used for the image acquisition and subsequent selection of the region of interest (ROI), whose size was determined by the confocal aperture, being approximately equal to one micron in the entire spectral range. To probe the absorption spectrum, the mirror was removed and the laser light stimulation mode of the CLSM was activated for the chosen ROI. This mode made it possible to project onto the input facet of the multimode optical fiber a preselected ROI only. The spectra were recorded by the fiber optic spectrometer connected to the CLSM by the multimode optical fiber. As a result, the setup allowed the measurement of the transmission spectra with a spectral resolution of up to 1 nm in the wavelength range of 400-800 nm. The spatial resolution of the measurements was determined by the numerical aperture of the receiving objective and the confocal aperture. The absorption was measured relative to the adjacent areas free from particles with the additional subtraction the background signal. The measurements were carried out using the objective lens UPLSAPO 40X NA 0.90. The experimental setup was controlled by Olympus FV10 ASW 1.7 software and Ocean Optics OceanView 1.6.7 software, and the processing of the data to obtain the spectra from the selected regions of interest was done using the same software.

**4.5. Laser heating of gold-filled spherulites**

For calibration the quartz cuvette with the sample was placed into the thermostat Luma 40/Horiba4 (Quantum Northwest, USA), and the fluorescence lifetime of Cy5 was recorded for the temperature varying from 20 to 55 ℃ using the custom-built fluorometer described in Ref.[40]. The dye excitation was performed by a 635-nm laser diode (IOS, Saint Petersburg, Russia) generating 40-ps pulses with a repetition rate of 10 MHz. The registration system which includes a photomultiplier (PMC-100, Becker&Hickl, Germany) and a single photon counter module (SPC-130EM, Becker&Hickl, Germany) was used to detect emission. The results are shown in Figure S6.

Aqueous solutions of gold-filled spherulites was heated with the "Gem" laser (LaserQuantum, Germany). The dye was excited by a 635-nm laser diode (IOS, Saint

Petersburg, Russia) generating 40-ps and 10-pJ pulses with a repetition rate of 10 MHz. A photomultiplier (PMC-100, Becker&Hickl, Germany) and a single photon counter module (SPC-130EM, Becker&Hickl, Germany) were used to detect emission. The heating laser was turned off during the fluorescence signal acquisition, which took 30 s.

The FLIM measurements were performed with the Microtime200 system (PicoQuant, Germany). The dye was excited at 640 nm with 40-ps and 1-pJ pulses at a repetition rate of 40 MHz, and the fluorescence signal was detected through the 655-725 nm bandpass filter. Heating at the wavelength 640 nm was performed by the laser of the Microtime200 system and at 785 nm by the Turnkey Raman 785nm Laser (Ocean Optics, USA).

**Acknowledgements**

Roman E. Noskov, Andrey Machnev, and Ivan I. Shishkin contributed equally. This research was partially supported by the Naomi Foundation through the Tel Aviv University GRTF Program, ERC StG 'In Motion' (802279), PAZY Foundation (Grant No. 01021248), Tel Aviv University Breakthrough Innovative Research Grant, the Ministry of Science, Technology and Space of Israel, the Russian Foundation for Basic Research (RFBR grant № 18-29-08046) — in the parts of the vaterite particle synthesis and gold nanoparticle loading into vaterite particles using the FIL method, and the Russian Science Foundation (grants № 19-13-00332 and № 20-45-08004 (absorption and FLIM measurements)). The work of E.A.S., A.V.G and A.A.E. was performed according to the Development program of the Interdisciplinary Scientific and Educational School of Lomonosov Moscow State University «Photonic and Quantum technologies. Digital medicine». R.E.N. is grateful to A. S. Shalin and E. A. Gurvitz for useful discussions. Andrey Machnev acknowledges the scholarship from "The Marian Gertner Institute for Medical Nanosystems". The authors acknowledge Dr. Gal Radovsky for help with high-resolution SEM as well as Thermo Fisher and Guillaume Amiard for help with FIB-STEM characterization of the spherulites.

# Supporting Information

**Golden Vaterite as a Mesoscopic Optical Metamaterial**

*Roman E. Noskov\*, Andrey Machnev, Ivan I. Shishkin, Marina V. Novoselova, Alexey V. Gayer, Alexander A. Ezhov, Evgeny A. Shirshin, Sergey V. German, Ivan D. Rukhlenko, Simon Fleming, Boris N. Khlebtsov, Dmitry A. Gorin, and Pavel Ginzburg*

# 1. Statistical analysis of SEM and STEM micrographs.

Processing of the SEM micrographs of individual particles and their clusters was first performed using image segmentation by Otsu's method[1]. This method allows the pixels belonging to vaterite spherulites and large clusters of gold to be filtered out from the background by using an intensity threshold of the histogram. However, the straightforward image thresholding did not permit detection of local intensity variations of just a few pixels in size, which are associated with individual Au NPs. To resolve this issue, we applied a Hessian feature detector to the original images[2,3]. This resulted in two sets of post-processed images, with the large-scale and small-scale features distinguished. Their sum yielded a mask where all the single spherulites and gold nanoparticles were recognized, as shown in Figure S1a. The surface coating factor was calculated as a fraction of the area occupied by Au NPs to the visible area of the spherulite. The resulting distributions are shown in Figure S1b. Fitting this data with the Gaussian probability density function allowed us to obtain the mean and the variance as functions of the number of loading cycles (Figure 1c).

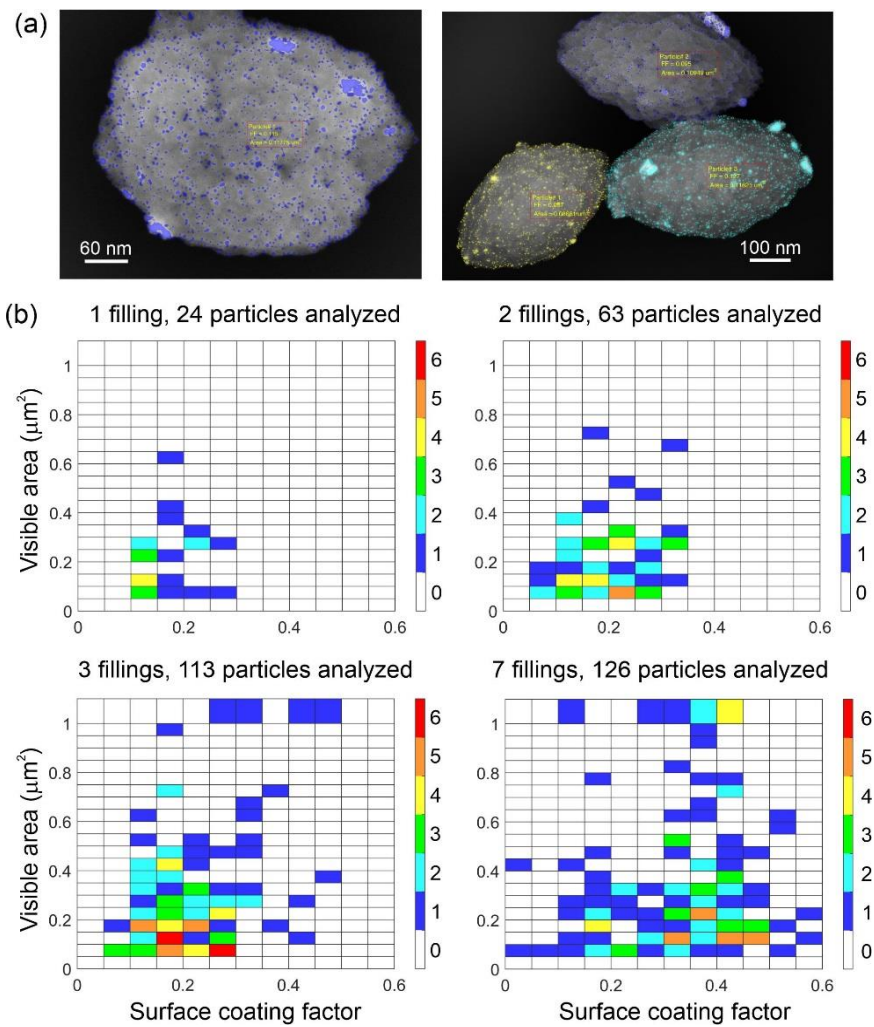

**Figure S1.** (a) Example of segmented SEM micrographs with colorized Au NPs and (b) number of vaterite spherulites (marked by colors in accordance with the color bar) versus the visible area and the surface coating factor of gold.

The volumetric filling factors were calculated by processing a set of STEM slices (z-stack) for the sample with 7 loading cycles. The cluster of spherulites was controllably etched by a focused ion beam gun with a step of 10 nm and imaged after each step by STEM with a spatial resolution of 1 nm. Owing to the presence of many Au NPs around the vaterite spherulites, the straightforward image thresholding for particle detection did not provide satisfying results. Instead, a sum of the image hessians on different scale levels was calculated for each individual image. By thresholding a normalized sum of image hessians, it became possible to distinguish the spherulites outlines, which were transformed into masks marking individual particles by the morphological closing of the resulting binary image.

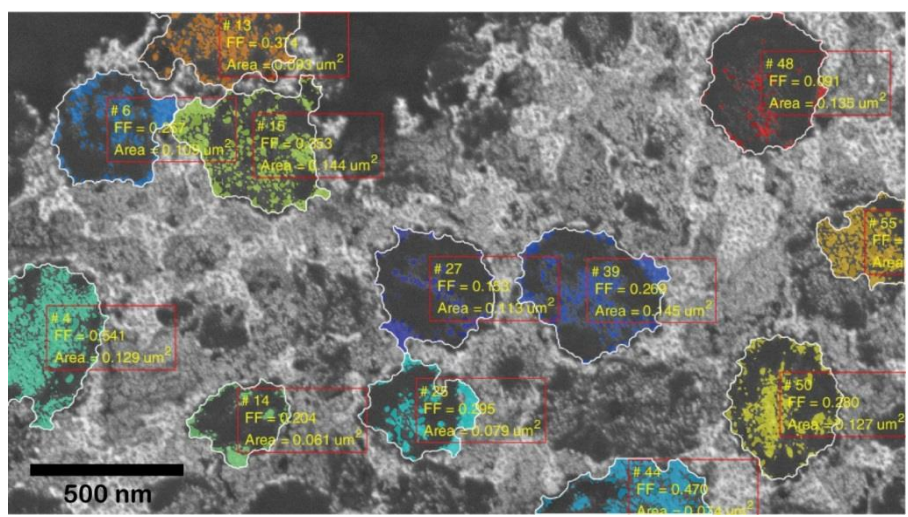

**Figure S2.** STEM image of a FIB slice with colorized Au NPs and spherulite outlines marked by white curves. FF means filling factor.

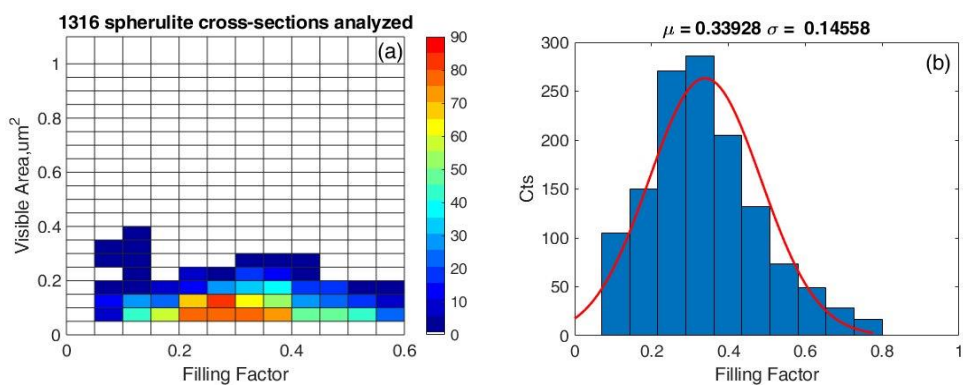

**Figure S3.** (a) Number of vaterite spherulite cross-sections (marked by colors as per the color bar) versus cross-section area and filling factor and (b) corresponding probability distribution function.

Pixels corresponding to Au NPs were identified by combining k-means segmented images and the thresholded sum of the image hessians. K-means segmentation was carried out for pixel intensity values. The total filling factor was calculated as the fraction of pixels corresponding to Au NPs of a given particle cross-section over the total area of this cross-section. An exemplary segmented image with total filling factors for individual spherulite cross-sections is given in Figure S2. The result of statistical data processing is presented in Figure S3. The mean total filling factor $f_0 = 0.34$ with the standard deviation $\sigma = 0.14$ were obtained.

Likewise, we also distinguished voxels corresponding to individual spherulites as well as Au NPs inside them by processing STEM z-stacks and calculated the volume filling factors shown in Figure 1c. Figure S4 shows examples of such 3D reconstructions. We obtained $f_0 = 0.28$ and $\sigma = 0.06$ by processing 10 spherulites which are in good agreement with the result of processing separate spherulite cross-sections.

Finally, we extracted the spatial distributions of Au NPs inside vaterite spherulites. To this end, we assumed that the gold concentration depends on the distance from the spherulite's center only, split them into many co-axial spherical layers, and calculated a local volumetric filling factor in each layer, shown in Figures 1f and S4. We found that these distributions are best fitted by a sum of two Gaussians as follows:

$$\rho(r) = A_1 \exp\left[-\frac{(r - \Delta r_1)^2}{2\delta_1^2}\right] + A_2 \exp\left[-\frac{(r - \Delta r_2)^2}{2\delta_2^2}\right],$$

where $A_{1,2}$, $\Delta r_{1,2}$, and $\delta_{1,2}$ are fitting parameters. Table S1 shows these fitting parameters for the distributions shown in Figures 1 and S4.

|  | Figure S4a | Figure S4b | Figure S4c | Figure S4d | Figure 1f |
|---|---|---|---|---|---|
| $A_1$ | 0.4 | 0.4 | 0.2 | 0.42 | 0.3 |
| $A_2$ | 0.38 | 0.46 | 0.28 | 0.49 | 0.42 |
| $\Delta r_1$ | 0.5 | 0.5 | 0.4 | 0.5 | 0.45 |
| $\Delta r_2$ | 1.02 | 1.03 | 1 | 1.05 | 1.02 |
| $\delta_1$ | 0.25 | 0.3 | 0.22 | 0.26 | 0.25 |
| $\delta_2$ | 0.15 | 0.17 | 0.22 | 0.2 | 0.21 |
| $f_0$ | 0.3 | 0.35 | 0.18 | 0.35 | 0.28 |

**Table S1.** Fitting parameters and the total filling factors for spatial distributions of Au NPs inside vaterite spherulites shown in Figures 1 and S4.

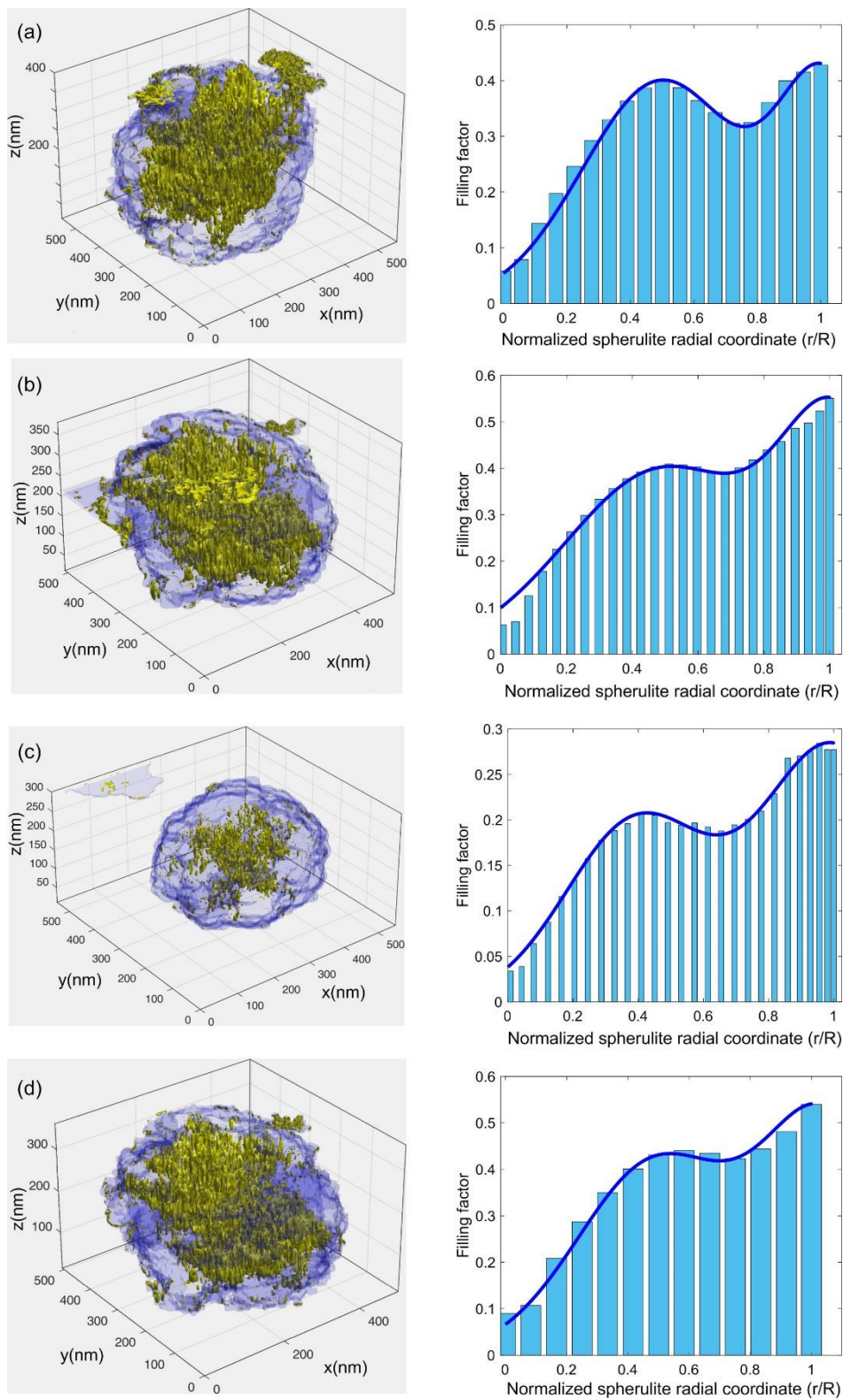

**Figure S4.** 3D reconstructions of vaterite spherulites along with corresponding spatial distributions of Au NPs inside them obtained by processing a z-stack of STEM images. The resolutions across the x, y, and z axes are 1, 1, and 10 nm. The distributions are fitted by a sum of two Gaussians with parameters given in Table S1.

## 2. Absorption spectroscopy setup

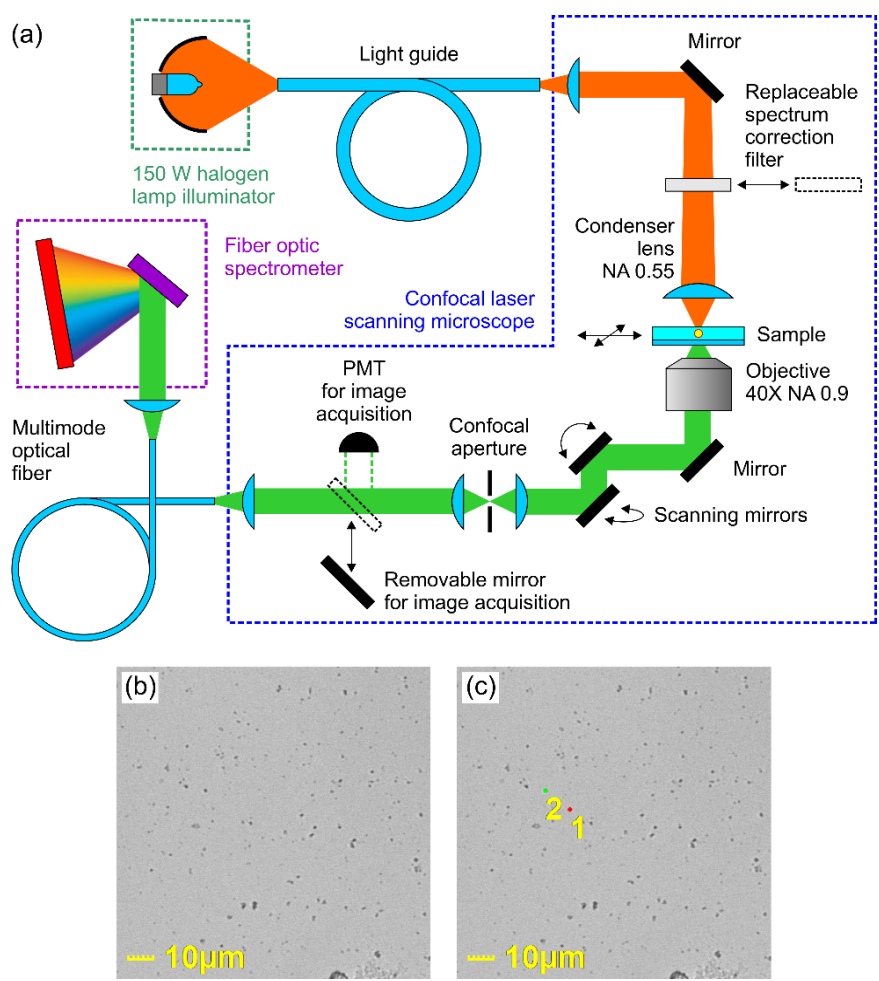

**Figure S5.** (a) Schematic of the setup for measurement of the absorption spectra for individual vaterite spherulites, (b) typical image obtained to select regions of interest (ROIs) and (c) ROIs in which (1) the reference spectrum and (2) the spectrum of the vaterite spherulite were measured.

## 3. Calibration for laser heating measurements

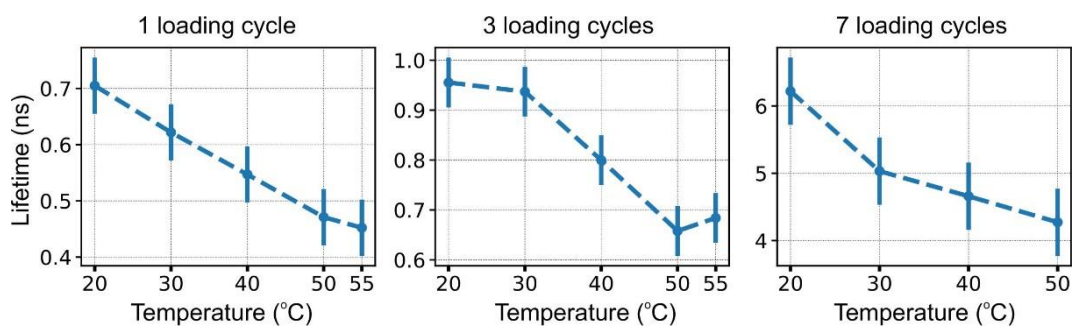

**Figure S6.** Calibration relationships between the temperature and the fluorescence lifetime for gold-filled vaterite spherulites obtained by a varying number of loading cycles.

## 4. Spherulite description in numerical model

A polycrystal vaterite spherulite can be represented by the so-called 'dumbbell' or 'heap of wheat' model[4] as a bundle of fibers tied together at the center and spread out at the ends. Each fiber consists of small (about 30 nm in size) positive uniaxial monocrystals with $\varepsilon_o \approx 2.4$ and $\varepsilon_{eo} \approx 2.72$, aligned to the tangents of the fiber. The distribution of the unity optical axes in the spherulite is well approximated by the family of co-focused hyperboles symmetrically rotated with respect to the $z$-axis[4,5] (Figure 1a). In what follows we assume that the focal position of the hyperboles is half of the radius of the sphere that fits submicron vaterite spherulites well.

Following Noskov $et\ al.$ [4], one can obtain the permittivity tensor of a spherulite by translating the local permittivity $\hat{\varepsilon}_{loc}$ of a uniaxial subunit into the global coordinate system related to the geometrical center of the spherulite as $\hat{\varepsilon}_{vat} = \widehat{M}_z(\varphi)\widehat{M}_y(\psi)\hat{\varepsilon}_{loc}\widehat{M}_y^{-1}(\psi)\widehat{M}_z^{-1}(\varphi)$, where

$$\hat{\varepsilon}_{loc} = \begin{pmatrix} \varepsilon_o & 0 & 0 \\ 0 & \varepsilon_o & 0 \\ 0 & 0 & \varepsilon_{eo} \end{pmatrix},$$

$\widehat{M}_y(\psi)$ and $\widehat{M}_z(\varphi)$ are the rotation matrixes around the $y$ and $z$ axes, $\varphi$ is the azimuthal angle, and $\psi$ is the angle between the $z$-axis and the tangent line at the local point of the hyperbola. This procedure results in the symmetric non-diagonal permittivity tensor

$$\hat{\varepsilon}_{vat} = \begin{pmatrix} \varepsilon_{xx} & \varepsilon_{xy} & \varepsilon_{xz} \\ \varepsilon_{xy} & \varepsilon_{yy} & \varepsilon_{yz} \\ \varepsilon_{xz} & \varepsilon_{yz} & \varepsilon_{zz} \end{pmatrix},$$

where

$$\varepsilon_{xx} = \varepsilon_o \left[ \cos^2 \varphi \left( \cos^2 \psi + \eta \sin^2 \psi \right) + \sin^2 \varphi \right],$$

$$\varepsilon_{yy} = \varepsilon_o \left[ \sin^2 \varphi \left( \cos^2 \psi + \eta \sin^2 \psi \right) + \cos^2 \varphi \right],$$

$$\varepsilon_{zz} = \varepsilon_o \left[ \sin^2 \psi + \eta \cos^2 \psi \right],$$

$$\varepsilon_{xy} = \varepsilon_{yx} = \varepsilon_o \sin \varphi \cos \varphi \left[ \cos^2 \psi + \eta \sin^2 \psi - 1 \right],$$

$$\varepsilon_{xz} = \varepsilon_{zx} = \varepsilon_o (\eta - 1) \cos \varphi \sin \psi \cos \psi,$$

$$\varepsilon_{yz} = \varepsilon_{zy} = \varepsilon_o (\eta - 1) \sin \varphi \sin \psi \cos \psi,$$

and $\eta = \varepsilon_{eo} / \varepsilon_o$. The equations of the hyperbola and its tangent make this model self-consistent,

$$\begin{cases} \cot \psi = \dfrac{b^2}{a^2} \tan \vartheta \\ r = \dfrac{ab}{\sqrt{\left( a^2 + b^2 \right) \sin^2 \vartheta - a^2}} \end{cases},$$

where $a$ and $b$ are the semi-major and semi-minor axes ($d^2 = a^2 + b^2$), and $r$ and $\vartheta$ are the radial distance from the spherulite center and the polar angle. Varying $a$ and $b$ from zero to infinity (i.e., filling the entire spherulite's cross-section with hyperbolic fibers) yields a parametric function $\psi(\vartheta, r)$, which makes all components of the spherulite's permittivity tensor coordinate-dependent.